\documentclass[epsf,12pt]{article}
\newcounter{multieqs}

\newcommand{\bfnabla}{\mbox{\boldmath $\nabla$}}
\newcommand{\bfvdd}{\mbox{\boldmath $\ddot v$}}
\newcommand{\bfvd}{\mbox{\boldmath $\dot v$}}
\newcommand{\bfF}{\mbox{\boldmath $F$}}

\newcommand{\bfx}{\mbox{\boldmath $x$}}
\newcommand{\bfj}{\mbox{\boldmath $j$}}

\newcommand{\bfA}{\mbox{\boldmath $A$}}

\newcommand{\bfv}{\mbox{\boldmath $v$}}
\newcommand{\bfV}{\mbox{\boldmath $V$}}
\newcommand{\bfr}{\mbox{\boldmath $r$}}
\newcommand{\bfrd}{\mbox{\boldmath $\dot r$}}

\usepackage{amssymb}
\usepackage{epsf}

\begin{document}

\begin{center}
{\large\bf THE SELF-FORCE OF A CHARGED PARTICLE IN
 CLASSICAL ELECTRODYNAMICS WITH A CUT-OFF}\\[0.7cm]
{\bf J. Frenkel} \ and \ {\bf R.B. Santos}\\
Instituto de F\'{\i}sica, Universidade de S\~{a}o Paulo\\[-0.2cm]
C.P. 66318, 05389-970 S\~{a}o Paulo, SP, Brazil    
\end{center}

\vspace{2cm}

\begin{center}
{\bf ABSTRACT}
\end{center}

We discuss, in the context of classical electrodynamics with a Lorentz
invariant cut-off at short distances, the self-force acting on a point
charged particle. It follows that the electromagnetic mass of the point
charge occurs in the equation of motion in a form consistent with
special relativity. We find that the exact equation of motion does not
exhibit runaway solutions or non-causal behavior, when the cut-off is
larger than half of the classical radius of the electron.



\vfill
\eject

\noindent{\bf I - INTRODUCTION}
\vspace{0.3cm}

The calculation of the self-force acting on a charged particle is a
long outstanding problem in electrodynamics since the days of Abraham
(1903) and Lorentz (1904), who derived for the first time the radiation
reaction force on an extended electron$^1$. Assuming that
the electron has a spherically symmetric rigid charge distribution of
radius $\, r_0\,$ in its instantaneous rest frame, they were able to
show that a particle subjected to an external force $\,{\bfF}_{\rm
ext}\,$ obeys the following equation of motion:
\begin{equation}
\frac{4}{3} \; \frac{U}{c^2} \, \bfvd - \frac{2}{3} \;
 \frac{e^2}{c^3} \, \bfvdd + \frac{2e^2}{3c^3} \,
  \sum^{\infty}_{n=2}\, \frac{(-1)^n}{n! c^n} \, \gamma_n \,
   \frac{d^{n+1} \, {\bfv}}{dt^{n+1}} = \bfF_{\rm ext}  
\end{equation} 
where $\, \bfvd\,$ is the acceleration of the particle and $\, U\,$
represents its electrostatic energy:
\begin{equation}
U \, = \, \frac{1}{2} \int d^3 x \int d^3 x' \; 
            \frac{\rho(\bfx)\rho(\bfx')}{|\bfx - \bfx'|} \ \ .
\end{equation} 
The constants $\, \gamma_n\,$ are proportional to $\, r_0^{n-1}\,$ and
characterize the way the charge is distributed within the particle.

The factor 4/3 in front of the electromagnetic mass $\, U/c^2\,$ in
equation~(1) is just one of the several well-known difficulties
involved in the Abraham-Lorentz theory of charged particles. To
overcome this problem, Poincar\'{e} (1905) suggested that the charged
particle could not be held together unless other attractive and
nonelectromagnetic forces were present. These Poincar\'{e} stresses
would add a mass $\, m_0\,$ to the electromagnetic mass of the
particle, so that the requirements of special relativity would apply
only to the physical, observed mass $\, m = m_0 + m_{\rm el}$.
However, there are two problems with that solution. Firstly, a charged
particle endowed with Poincar\'{e} stresses would be unstable under
deformations of shape which might occur when the particle is acted by
external or self-electromagnetic forces$^1$. Secondly,
classical electrodynamics is a covariant theory by itself, so one
expects that a correct calculation should not violate the requirements
of Lorentz covariance. The current point of view is that the
electromagnetic energy-momentum used by Abraham, Lorentz and
Poincar\'{e} is not a covariant quantity, so that when the covariance
condition is taken properly into account it should furnish the expected
factor of unity.$^{2-6}$

Anyway, the extended electron theory is not compatible with the
experimental facts, which indicate that the electron may be considered
as a point particle at least up to distances of order of
$\,10^{-16}\,$cm.$^7$ \ In the point charge limit, all the
structure-dependent constants $\, \gamma_n\,$ in equation~(1) go to
zero, but then the electromagnetic mass $\, m_{\rm el}\,$ diverges as
$\, 1/r_0\,$ when $\, r_0 \rightarrow 0$, so that this limit is not
meaningful in the Maxwell theory. One may assume that the terms
involving the $\, \gamma_n\,$ factors in equation~(1) could be
disregarded when $\, r_0\,$ is very small, provided the changes in the
motion of the particle which occur during short time intervals of order
$\, r_0/c\,$ are negligible. One then obtains the Abraham-Lorentz
equation of motion:
\begin{equation}
\left( m_0 + m_{\rm el} \right) \, \bfvd - \frac{2}{3} \; 
   \frac{e^2}{c^3} \, \bfvdd = \bfF_{\rm ext}
\end{equation} 
where we have added a mechanical nonelectromagnetic mass $\, m_0$. As
remarked by Feyn\-man$^8$, one would be in trouble only if
the energy changes were also infinite. Unfortunately, this is the case:
even if we renormalize the mass, keeping $\, m = m_0 + m_{\rm el}\,$
fixed as $\, r_0 \rightarrow 0$, the solution of equation~(3) when $\,
\bfF = 0\,$ would have an exponentially growing acceleration:
\begin{equation}
\bfvd (t) = \bfvd (0) \, \exp(t/\tau)
\end{equation} 
where $\, \tau = 2e^2/3mc^3$. This is called a runaway solution of the
Abraham-Lorentz equa\-tion.$^{3-6,8-10}$ \ When there is an external
force acting on the charged particle, the runaway solution still
persists: 
\begin{equation}
\bfvd (t) = \left[ \bfvd (0) - \frac{1}{m\tau} \int^t_0 dt'\, \exp
  (- t'/\tau) \, \bfF_{\rm ext} (t') \right] \exp(t/\tau) \ ,
\end{equation} 
unless we impose, following Dirac$^9$, the very peculiar 
initial condition:
\begin{equation}
\bfvd (0) = \frac{1}{m\tau} \int^\infty_0 dt'\, \exp(- t'/\tau) \,
   \bfF_{\rm ext}(t') \ .
\end{equation}
But in this case, the acceleration $\, \bfvd (t)\,$ would depend on the
force $\, \bfF_{\rm ext} (t+t')\,$ at times greater than $\, t$. This
non-causal effect, which is more pronunciated during times $\, t'\,$ of
order $\, \tau\,$ is called preacceleration.

The above behavior indicates that the assumption about the neglect of
the $\, \gamma_n\,$ factors in this regime may be inconsistent, since
during short time intervals of order $\,\tau\,$ the changes in the
motion of the particle appear to be important.  On the other hand it is
well known that due to quantum effects, classical electrodynamics
cannot remain valid at such small distances and time intervals when the
runaways and the preacceleration effects are relevant.  Thus, it is
possible that a modification of the laws of electrodynamics at short
distances might lead to a regularized, causal and runaway-free theory.
In fact, because of the existence of a cut-off in such a theory, we
shall show that the $\, \gamma_n \,$ terms are nonvanishing in the
point particle limit, being essential for the suppression of the
unphysical runaway solutions. 

Some time ago, Coleman$^{11}$, treating the electron as a point charge
from the very beginning, introduced a cut-off in Maxwell's
electrodynamics. This enabled him to derive unambiguously the
relativistic equation of motion which reduces to equation~(3) in the
nonrelativistic limit, called the Lorentz-Dirac equation.$^9$ \ In his
work, the cut-off was merely a computational device, whose effects were
disregarded at the end of the calculation. A few years later, Moniz and
Sharp$^{12,13,14}$ have shown in the context of a quantum theory of the
electron, that the interaction of the point electron with his own
electromagnetic field induces effectively a natural cut-off of order of
the electron's Compton wavelength $\, \lambda = \hbar/mc$. This may
arise in consequence of the creation of virtual electron-positron pairs
in the neighbourhood of the point electron, which effectively
spread-out its charge distribution. Subsequently, these and other
related aspects have been further investigated by several
authors.$^{15-20}$

Based on these facts, we believe that a possible way to remove the
divergences, runaway solutions or noncausal behavior from classical
electrodynamics is by the introduction of a gauge and Lorentz invariant
cut-off at short distances in the Maxwell theory. We shall use such a
cut-off at the threshold of the classical regime, which allows for the
existence of a finite and well defined point particle limit. One of the
authors$^{21}$ has recently shown how calculating the electromagnetic
mass in this framework solves the 4/3 problem of the classical theory.
In the regularized classical electrodynamics one finds the correct
factor of unity in the point charge limit as well as a finite
electromagnetic mass $\, m_{\rm el} = e^2/2\ell\, c^2$, where $\, \ell
\,$ is the cut-off. When $\, \ell = \hbar/mc\,$ this becomes, apart
from a logarithmic factor, of same order as the electromagnetic mass
found in quantum electrodynamics.$^{22}$

The approach we follow involves adding a new term to the Maxwell
Lagrangian, which leads to an effective Lagrangian for classical
electrodynamics that takes into account the relevant effects from the
quantum theory. The form of the new term can be restricted by a few
reasonable and simple properties, which leave the Maxwell theory as
unaltered as possible: (a)~The Lagrangian must be gauge and Lorentz
invariant. \ (b)~It should yield local field equations which are still
linear in the field quantities. The simplest possibility that includes
a cut-off $\, \ell\,$ leads to a Lagrangian containing second order
derivatives of the electromagnetic potentials $\, A_\alpha = (\bfA,
i\phi)$:
\begin{equation}
{\cal L}(\ell\,) \; = \; - \, \frac{1}{16\pi} \, F_{\alpha\beta} \,
   F_{\alpha\beta} \, - \, \frac{\ell\,^2}{8\pi} \;
    \partial_\beta\, F_{\alpha\beta} \, \partial_\gamma\, 
     F_{\alpha\gamma} + \frac{1}{c} \, j_\alpha \, A_\alpha \ , 
\end{equation}
where $\, F_{\alpha\beta} = \partial_\alpha \, A_\beta - \partial_\beta
\, A_\alpha\,$ is the usual electromagnetic field tensor and $\,
j_\alpha = (\bfj, ic\rho)\,$ is the conserved four-current. At
distances much larger than the cut-off, the fields described by
equation~(7) become essentially equivalent to the fields governed by
the usual Maxwell theory. Such a modification of classical
electrodynamics was proposed a long time ago by Podolsky and
others.$^{23}$  

In section II we present the calculation of the self-force that acts on
a point charged particle, within the framework of the generalized
Maxwell theory described by the Lagrangian~(7). We evaluate all the
terms involving higher order derivatives of the velocity which appear
in the exact equation of motion of a point charged particle. The
contributions of the higher order terms can be summed in closed form,
from which the absence of runaway behavior follows in the case when the
cut-off is larger than half of the classical radius of the electron.
Furthermore, we find that in this case the solutions are consistent
with the principle of causality. Similar conclusions may be obtained
from the exact equation of motion of a relativistic point charged
particle, which is discussed in the last section.

\vspace{0.8cm}

\begin{list}{}{\setlength{\leftmargin}{7mm}\labelwidth2.5cm
 \itemsep0pt
\parsep0pt} 

\item[{\bf II.}] {\bf EVALUATION OF THE SELF-FORCE OF A POINT CHARGED 
PARTICLE}
\end{list}
\vspace{0.2cm}

The Lagrangian $\, {\cal L} (\ell\,)\,$ leads to the following linear
partial differential equations: 
\begin{equation}
\left( 1 - \ell\,^2 \square \right) \square A_\alpha \; = \; - \, 
    \frac{4\pi}{c} \, j_\alpha
\end{equation} 
where we used the Lorentz gauge $\,\partial_\alpha\,A_\alpha = 0$. To
determine these potentials, it is useful to find the retarded Green
function for the equation
\begin{equation}
\left( 1 - \ell\,^2 \square \right) \square  G( \bfx - \bfx', 
   t - t', \ell \,) \, = \, - \, 4\pi \, \delta 
    (\bfx - \bfx') \,\delta(t - t') 
\end{equation} 
which is subjected to the causality condition that $\, G=0\,$ for $\, t
< t'$.  In that way the solution of equation~(8) will be:
\begin{equation}
A_\alpha \left( \bfx, t,\ell\, \right) \, = \, \frac{1}{c} \int 
  d^3x'\, dt'\, G \left( \bfx - \bfx', t - t', \ell \, \right) j_\alpha
    \left( \bfx', t' \right)  \ .
\end{equation} 
Following the procedure described in reference~21, we
arrive at 
\begin{equation}
G(R,T,\ell \,) \, = \, 
    \frac{c\,\theta (T-R/c)}{\ell\sqrt{c^2T^2 -R^2}}\,
   J_1 \left( \frac{\sqrt{c^2T^2 - R^2}}{\ell} \,\right) \ ,
\end{equation} 
where $\, R = |\bfx - \bfx'|, \; T = t - t'\,$ and $\, J_1\,$ is the
Bessel function of order one. The self-force may be shown to be:
\begin{equation}
\bfF_s\,(t) = - \int d^3x \,\rho(\bfx, t) \left[ \bfnabla \, \phi(\bfx,  
   t) + \frac{1}{c} \; \frac{\partial \bfA}{\partial t} \, 
    (\bfx, t) \right]  \ .
\end{equation} 
Now, instead of performing a series expansion in powers of $\, R/c$,
for $\, R/c\,$ small, of the retarded Green function $\,
G(R,T,\ell\,)$, we will rather take the point particle limit in
equation~(11), 
\begin{equation}
G(0,T,\ell\,) = \frac{\theta(T)}{\ell\, T} \; J_1 (c\,T/\ell\,) \ , 
\end{equation}
which allows us to express the self-force in a closed form given by the
expression:
\begin{equation}
{\bfF}_s(t) \, = \, \frac{e^2}{c^2} \int^\infty_0
    dT  \; \frac{dG(0,T,\ell\,)}{dT} \,  
     \left[\frac{\bfr(t) - \bfr(t-T)}{T} - \bfv
      (t-T)\right] \ ,
\end{equation} 
where $\, \bfr \,$ is the coordinate of the particle and $\, \bfv(t) =
\bfrd (t)\,$ is its velocity.

Unlike the Abraham-Lorentz equation, expression~(14) does not involve
any explicit second order derivatives of the velocity with respect to
time. For that reason, the exact equation of motion of the particle, 
\begin{equation}
m_0\bfvd - {\bfF}_s(t) = {\bfF}_{\rm ext}
\end{equation} 
has substantially different properties from those of the Abraham-Lorentz
equation. In particular, the homogeneous solutions of equation~(15) do
not display runaway behavior when $\, m_0\,$ is non-negative.

To see this we assume an ansatz of the form $\, \bfr =\bfr_0
\,\exp(\eta t)$. There will be no runaway solutions if the real part of
$\, \eta\,$ is negative or vanishes. Then, the possible solutions of
equation~(15) in the absence of external forces are determined by the
condition that
\begin{equation}
m_0 \, \eta^2 + \frac{e^2}{c^2\ell} \int^\infty_0 dT \left[ 
  \exp(-\eta T) - 1 + \eta T \, \exp(-\eta T)  
   \right] \, \frac{1}{T} \; \frac{d}{dT} \, \left[ 
    \frac{J_1(c\,T/\ell\,)}{T} \right] = 0 \ .
\end{equation} 
The $T$-integration may be performed,$^{24}$ \ giving
\begin{equation}
\left( \eta^2 + \frac{c^2}{\ell\,^2} \right)^{1/2} \,
   \left( 2\eta^2 \, - \, \frac{c^2}{\ell\,^2} \right) = 
    2\eta^3 \,- \, \frac{c^3}{\ell\,^3} \, - \, 
     \frac{3m_0 \, c^3}{e^2} \, \eta^2 \ .
\end{equation} 

Squaring both sides and noticing that $\, \eta =0\,$ is a doubly
degenerate solution, we may rewrite (17) as the cubic equation:
\begin{equation}
m_0 \, \eta^3 \, - \, \frac{3}{4} \, \frac{m_0^2 \, c^3}{e^2} \, \eta^2
  + \frac{1}{3} \, \frac{e^2}{\ell^3} \, \eta \, - \, \frac{1}{4} \,
    \frac{c\,e^2}{\,\ell\,^4}\, - \, \frac{1}{2} \, \frac{m_0 \,
		  c^3}{\,\ell\,^3} \, = \, 0 \ . 
\end{equation} 

The solutions of (17) are a subset of those determined by the cubic
equation~(18). The solutions of such an equation are well known and it
can be verified that its complex conjugate roots do not satisfy the
original equation~(17). Therefore, we are allowed to write the
solutions of (17) as $\, \eta = cx/\ell$, where $\, x\,$ is a real
quantity satisfying equation~(17) in the form
\begin{equation}
\left( 1 + x^2 \right)^{1/2} \,(2x^2-1) = 2x^3 -1-px^2 \ ,
\end{equation} 
where $\, p = 3 m_0\ell c^2/e^2\,$ is a dimensionless real parameter.
Note that the sign of the mechanical mass $\, m_0\,$ determines that of
$\, p$.  

Apart from the trivial solution $\, x=0$, we must distinguish three
cases in order to find the other solutions:

\begin{list}{}{\setlength{\leftmargin}{7mm}\labelwidth2.5cm
 \itemsep0pt
\parsep0pt} 
\item[(i)]~$p=0$. \ A very simple analysis shows that the left hand side of
(19) is always larger than the right hand side, except when $\, x=0$.
Hence, there are no additional solutions when $\, p=0$.

\item[(ii)]~$p > 0$. \ The above argument holds still stronger in this case.
Thus, we cannot get extra solutions in this case either. 

\item[(iii)]~$p < 0$. There is a continuous set of solutions $\, x = x(p)$. To
see that, consider the inverse relation $\, p = p(x)\,$ which,
according to (19), is given by

\end{list}
\begin{equation}
p = 2x - \frac{1}{x^2} + \left( 1 + x^2 \right)^{1/2} \,
     \left( \frac{1}{x^2} - 2 \right) \ .
\end{equation}
When $\, x \gg 1$, $\, p\,$ approaches zero as $ \, - 1/x^2$, while for
$\, x \ll 1$, $\, p\,$ behaves approximately as $\, - 3/2 + 2x$. A plot
of the graph of $\, p\,$ versus $\, x\,$ helps us to grasp these
features (see figure~1).


\begin{figure}[htb]
   \hspace{.1\textwidth}
   \vbox{\epsfxsize=.7\textwidth
    \epsfbox{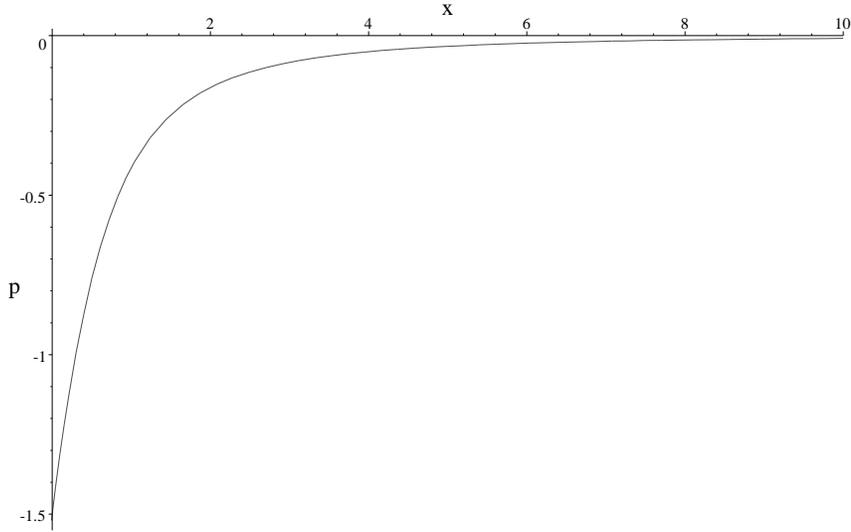}}
   \label{Fig1}
\caption{Behavior of the parameter $\, p\,$ as a function of the root
$\, x$.}
  \end{figure}   
%
%

\bigskip

Since $\, m = m_0 + m_{\rm el}$, and the electromagnetic mass $\,
m_{\rm el}\,$ is given in the regularized electrodynamics for a point
particle by$^{21}$
\begin{equation}
m_{\rm el} = \frac{e^2}{2\ell \, c^2} \ \ ,
\end{equation}  
it is possible to express $\, p\,$ in terms of the cut-off $\, \ell\,$
and the classical radius of the electron $\, r_0 = e^2/mc^2\,$ as:
\begin{equation}
p=\frac{3m_0\ell c^2}{e^2} = -\frac{3}{2} + \frac{3\ell}
      {r_0} \ .
\end{equation}
This shows that $\, p\,$ is necessarily larger than $\, - 3/2$, for
both $\, \ell \,$ and $\, r_0\,$ are positive constants. Therefore, we
are led to the conclusion that $\, x\,$ must be restricted to positive
values. Consequently, if $\, p < 0$, and hence $\, m_0 < 0$, $\,
\eta\,$ is real and positive and runaway motion takes place.

We see that runaway solutions can be presented if, and only if, $\, m_0
< 0$. Expression~(22) shows that this is possible provided $\, \ell\,$
is smaller than half the classical radius of the electron. It is
interesting to examine these solutions in the limit $\, \ell
\rightarrow 0$. In this case, we would have $\, x \simeq 3\ell/2r_0$.
Thus, the homogeneous solution of the exact equation of motion (15) may
be written in the limit $\, \ell \rightarrow 0\,$ as
\begin{equation}
\bfvd (t) = \bfvd_0\exp(\eta t) = \bfvd_0\, \exp[cx(p)t/\ell]
     \simeq \bfvd_0\exp(3ct/2r_0)
\end{equation} 
which is identical to the homogeneous solution (4) of the
Abraham-Lorentz equation of motion (3).

Let us finally examine the inhomogeneous solution of the regularized
equation of motion (15), corresponding to a nonrelativistic motion of
the particle, subject to a time-dependent external force. This solution
is easily obtained after introducing the Fourier transforms
\begin{equation}
\bfr(t) = \frac{1}{2\pi} \int^{+\infty}_{-\infty} d\omega \, 
    \exp(-i\omega t) \,\bfr(\omega) \ ;\qquad 
{\bfF}_{\rm ext} (t) = \frac{1}{2\pi} \int^{+\infty}_{-\infty} d\omega 
  \, \exp(-i\omega t)\, {\bfF}_{\rm ext} (\omega) \ .
\end{equation}

Using the expression (14) we arrive at the relation
\begin{eqnarray}
\bfr(\omega) & = & \left\{m_0(- i\omega)^2+
  \frac{e^2}{c^2\ell} \int^\infty_0 dT \left[ \exp(i\omega T) - 1
   + \right.\right. \nonumber \\[0.4cm]
& + & \left.\left. 
    (- i\omega T) \exp(i\omega T)\right] \frac{1}{T} \; \frac{d}{dT} 
     \left( \frac{J_1(cT/\ell)}{T}\right) \right\}^{-1} \,
            {\bfF}_{\rm ext}(\omega) \ .
\end{eqnarray}
Standard techniques, such as the convolution theorem, allows us to
write the inhomogeneous solution in the form 
\begin{equation}
\bfvd(t) = \int^{+\infty}_{-\infty} dt' \, 
  {\cal{G}}(t-t') {\bfF}_{\rm ext}(t') \ ,
\end{equation}
where the Green function $\,{\cal{G}}(t-t')\,$ is given by:
\begin{eqnarray}
&& {\cal{G}} (t-t') = \frac{1}{2\pi} \int^{+\infty}_{-\infty}  
    \exp(-i\omega(t-t'))(- i\omega)^2\,
      \left\{m_0(- i\omega)^2+ \frac{e^2}{c^2\ell} \int^\infty_0 dT 
        \times \right. \nonumber\\[0.4cm]
&& \left. \times  \left[ \exp(i\omega T) -1
   + (- i\omega T) \exp(i\omega T)\right] \frac{1}{T} \; \frac{d}{dT} 
     \left( \frac{J_1(cT/\ell)}{T}\right) \right\}^{-1} \, d\omega
\ .
\end{eqnarray}

The charge will move in a causal way if the acceleration at time $\,
t\,$ depends only upon the force field at times earlier than $\, t$.
That such a behavior can be ensured by the retarded Green function,
which is characterized by the fact that its singularities lie in the
lower half of the complex $\omega$-plane,$^{4,20}\,$ is well known.
Substituting $\, \eta\,$ for $\, (- i\omega)$, this property requires
that all zeros of the expression in curly brackets in equation~(27)
must be situated in the left half of the complex $\eta$-plane. But this
condition is identical to that given by equation~(16) in connection
with the absence of runaway solutions. It then follows from our
previous analysis that no improper solution is present, when the
cut-off is larger than half the classical radius of the electron.

One important aspect to note is that the removal of runaways and of
preacceleration is intimately related to keeping the higher order terms
in the expansion of the self-force.$^{21}$ \ In our framework, these
higher order terms do not vanish in the point particle limit because of
the existence of the cut-off $\, \ell$. Then, the self-force may be
written as
\begin{equation}
\bfF_s(t) = \sum^\infty_{n=0} \, \bfF^n_s(t) \ ,
\end{equation}
where
\begin{equation}
{\bfF}^n_s \, = \, \frac{b_n \, e^2}{n! \, c^{n+2}} \; \ell\,^{n-1} \; 
    \frac{d^{n+1}\, \bfv}{dt^{n+1}} \ \ ,
\end{equation} 
and the constants $\, b_n\,$ may be determined using the techniques
described in 21. We only cite the results. One finds that
$\, b_0 = 1/2, \; b_1 = - 2/3 \,$ and
\begin{equation}
b_n = \frac{(-1)^{n/2}\,(n+1)}{(n-1)(n+2)} \, 
        [(n-1)!!]^2
\end{equation} 
when $\, n \geq 2\,$ is even and $\, b_n = 0\,$ otherwise.

\noindent We see from the above equations that the factor $\, 2b_0$,
which multiplies the electromagnetic mass $\, e^2/2c^2\ell$, has the
correct value of unity which is consistent with special relativity.

\pagebreak

\noindent{\bf III. DISCUSSION}
\vspace{0.3cm}

A relativistic generalization of equation~(15) must have the form
\begin{equation}
m_0\,\frac{dv_\mu}{d\tau} - F_\mu= F_\mu^{\rm ext}
\end{equation}
where $\,\tau \,$ is the particle's proper time and $\, v_\mu\,$ is its
four-velocity:
\begin{equation}
v_\mu = \left( 1 - v^2/c^2 \right)^{-1/2} \, [ \bfv, ic] = 
       [ \gamma \bfv, i\gamma c ] \ .
\end{equation}
Here, $\, F_\mu\,$ represents the covariant generalization of the
self-force $\, \bfF_s$, which acts in the instantaneous rest frame of
the charged particle. Using the Lorentz transformation properties of
the four-vector $\, F_\mu$, together with the general constraint that
$\, F_\mu v_\mu =0$, we obtain for $\, F_\mu\,$ the expression
\begin{equation}
F_\mu = \left[\bfF_s + (\gamma-1)\,\frac{\bfv\cdot\bfF_s}
     {v^2}\,\bfv\, , \; i\gamma\,\frac{\bfv\cdot\bfF_s}
       {c}\right] \ .
\end{equation}
The self-force $\, F_\mu\,$ can be expanded in a power series involving
higher-order derivatives of the four-velocity with respect to the
proper time. This may be done conveniently using the relation
\begin{equation}
F_\mu =\frac{e}{c}\,F_{\mu\nu}\,v_\nu = \frac{e}{c}
      \, (\partial_\mu \,A_\nu - \partial_\nu  \,A_\mu)v_\nu \ ,
\end{equation}
where the self-potentials $\, A_\alpha\,$ given by equation~(10) must
be evaluated at the position of the point charge. The actual
calculation is rather involved and can be carried out along the lines
indicated in reference~11. The result is
\begin{equation}
F_\mu = -\frac{e^2}{2\ell c^2} \,\frac{dv_\mu}{d\tau} +
     \frac{2e^2}{3c^3}\left[\frac{d^2v_\mu}{d\tau^2} - 
       \frac{1}{c^2}\,\frac{dv_\nu}{d\tau}\,
         \frac{dv_\nu}{d\tau} \,v_\mu\right] +
           \sum^\infty_{n=2}\,\frac{b_n \, e^2}{n!c^{n+2}}
             \;\ell^{n-1} 
            \,V^n_\mu \  ,
\end{equation}
where $\,e^2/2\ell c^2\,$ is the electromagnetic mass of a point
particle and $\,b_n\,$ are the constants given in eq.~(30).  The
four-vector $\,V^n_\mu\,$ may be expressed in terms of the proper time
derivatives of the four-velocity, $\,v_\mu^{(n)}=
d^n\,v_\mu/d\tau^n\,$, as follows:
\pagebreak
\begin{eqnarray}
&& V^n_\mu=\frac{1}{(n+1)^2}\,\left[v_\mu^{(n+1)} +
    \frac{v_\nu\,v_\nu^{(n+1)}}{c^2}\,v_\mu
      \right]+ \nonumber\\[0.4cm]
&& +\,\frac{n+2}{n+1} \,\frac{v_\nu}{c^2}\,\sum^n_{k=1}\,
   \left(\begin{array}{c}
      n\\[-0.2cm]
         k\end{array}\right) \,\frac{1}{n+2-k}
         \left[v_\nu^{(n+2-k)}\,v_\mu^{(k-1)}- 
             (\mu \leftrightarrow \nu)\right] \ .
\end{eqnarray}
In the nonrelativistic regime the three-vector part of $\, V^n_\mu$,
namely $ \, \bfV^n$, is practically equal to $\, d^{n+1}\bfv/dt^{n+1}$,
in accordance with the result given by equation (29).

In the limit $\, \ell \rightarrow 0\,$ the self-force four-vector  
becomes 
\begin{equation}
F_\mu = - m_{\rm el} \; \frac{dv_\mu}{d\tau} +
     \frac{2e^2}{3c^3}\left[\frac{d^2v_\mu}{d\tau^2} - 
       \frac{1}{c^2}\,\frac{dv_\nu}{d\tau}\,
         \frac{dv_\nu}{d\tau} \,v_\mu\right] 
\end{equation} 
with a diverging electromagnetic mass $\, m_{\rm el} = e^2/2\ell\,
c^2$. In this limit, the exact equation of motion (31) reduces to the
Lorentz-Dirac equation with the physical mass $\ m= m_0 + e^2/2\ell \,
c^2$, 
\begin{equation}
m \, \frac{dv_\mu}{d\tau} -
     \frac{2e^2}{3c^3}\left[\frac{d^2v_\mu}{d\tau^2} - 
       \frac{1}{c^2}\,\frac{dv_\nu}{d\tau}\,
         \frac{dv_\nu}{d\tau} \,v_\mu\right] = F_\mu^{\rm ext} \ .
\end{equation} 
The Lorentz-Dirac equation is known to exhibit the familiar maladies of
runaway solutions and noncausal behavior. These problems may be
ascribed to the appearance of a negative bare mass $\, m_0\,$ to
counterbalance the diverging electromagnetic mass, in order to furnish
the observed, finite mass $\, m\,$ to the point charged particle.

The relativistic equation of motion (31), with $\, F_\mu\,$ given by
the exact expression (35), predicts in general the same kind of
behavior as that described by its nonrelativistic counterpart~(15). To
understand this feature, we remark that when $\, \ell < r_0/2$, the
mechanical mass $\, m_0\,$ must be negative in order to ensure the
observed value of the physical mass $\, m$. Then, a runaway behavior is
consistent with the conservation of energy, which is the sum of the
particle kinetic energy $(\gamma - 1)m_0c^2\,$ and the positive
electromagnetic field energy. The kinetic energy of a negative
mechanical mass, which is negative and decreasing with the increasing
velocity of the particle, can compensate the increase of the field
energy, maintaining an overall constant energy.

On the other hand, when $\, \ell\,$ is larger than half the classical
radius of the electron, $\, m_0\,$ is positive and the particle cannot
undergo a runaway motion. Such a motion would violate the conservation
of energy, since it would increase the particle positive mechanical
energy as well as the positive electromagnetic field energy. Therefore,
we conclude that if the quantum processes induce in the classical
regime an effective cut-off of order of the Compton wavelength of the
electron, then the exact equation of motion of a point charged particle
will admit only physical solutions.  

\vspace{1cm}

\noindent{\bf ACKNOWLEDGMENTS}

J.F. would like to thank Prof. J.C. Taylor for helpful discussions and
CNPq (Brasil) for a grant. R.B.S. is grateful to FAPESP (Brasil) for
financial support.

\pagebreak

\noindent{\bf REFERENCES}

\begin{list}{}{\setlength{\leftmargin}{5mm}\labelwidth2.5cm
\itemsep0pt \parsep0pt}

\item[1.] H.A. Lorentz, {\em The Theory of Electrons\/}, 2nd ed., Dover
publications, New York (1952).

\item[2.] B. Kwal, {\sl J. Phys. Radium\/} {\bf 10}, 103 (1949).

\item[3.]  F. Rohrlich, {\em Classical Charged Particles\/},
Addison-Wesley, Reading, MA (1965).

\item[4.] J.D. Jackson, {\em Classical Electrodynamics\/}, 2nd ed.,
Wiley, New York (1975).

\item[5.] Erber, {\sl Fortsch. d. Phys.\/} {\bf 9}, 343 (1961). 

\item[6.] P. Pearle, {\sl ``Classical Electron Models''}, in {\em
Electromagnetism: Paths to Research\/}, ed. D. Teplitz, Plenum Press,
New York (1982).

\item[7.] D.H. Perkins, {\em Introduction to High Energy Physics\/},
Addison-Wesley, Reading, MA (1987). 

\item[8.] R.P. Feynman, R.B. Leighton, M. Sands, {\em The Feynman
Lectures on Physics\/}, Addison-Wesley, Reading, MA (1964).

\item[9.] P.A.M. Dirac, {\sl Proc. Roy. Soc.\/} {\bf 167A}, 148 (1938).

\item[10.] P.W. Milonni, {\em The Quantum Vacuum: An Introduction to
Quantum Electrodynamics\/}, Academic, Boston (1994).

\item[11.] S. Coleman, {\sl ``Classical Electron Theory from a Modern
Standpoint''\/} (1960), reprinted in {\em Electromagnetism: Paths to
Research\/}, ed. D. Teplitz, Plenum Press, New York (1982).

\item[12.] E.J. Moniz and D.H. Sharp, {\sl Phys. Rev.\/} D{\bf 10},
1133 (1974).

\item[13.] E.J. Moniz and D.H. Sharp, {\sl Phys. Rev.\/} D{\bf 15},
2850 (1977).

\item[14.] H. Levine, E.J. Moniz and D.H. Sharp, {\sl Am. J.  Phys.\/}
{\bf 45}, 75 (1977).

\item[15.] H. Grotch and E. Kazes, {\sl Phys. Rev.\/} D{\bf 16}, 3605
(1977).

\item[16.]  D.H. Sharp, {\sl ``Radiation Reaction in Nonrelativistic
Quantum Theory''}, in {\em Foundations of Radiation Theory and Quantum
Electrodynamics\/}, A.O. Barut (ed.), Plenum, New York (1980).

\item[17.] H. Grotch and E. Kazes, {\sl ``Heisenberg Equation of Motion
Calculation of the Electron Self-Mass in Nonrelativistic Quantum
Electrodynamics''}, in {\em Foundations of Radiation Theory and Quantum
Electrodynamics\/}, A.O. Barut (ed.), Plenum, New York (1980).

\item[18.] F. Rohrlich, {\sl ``Fundamental Physical Problems of Quantum
Electrodynamics''}, in {\em Foundations of Radiation Theory and Quantum
Electrodynamics\/}, A.O. Barut (ed.), Plenum, New York (1980).

\item[19.] H. Grotch, E. Kazes, F. Rohrlich and D.H. Sharp, {\sl Acta
Phys. Austriaca\/} {\bf 54}, 31 (1982).

\item[20.] F.E. Low, ``Runaway electrons in relativistic spin 1/2
quantum electrodynamics'', hep-th/9702151 (1997).

\item[21.] J. Frenkel, {\sl Phys. Rev.\/} E{\bf 54}, 5859 (1996).

\item[22.] R.P. Feynman, {\em Quantum Electrodynamics\/}, W.A.
Benjamin, New York (1961).

\item[23.] B. Podolsky and P. Schwed, {\sl Rev. Mod. Phys.\/} {\bf 20},
40 (1948).

\item[24.] I.S. Gradshteyn and I.M. Ryzhik, {\em Tables of Integrals,
Series and Products\/}, Academic Press, New York (1980).

\end{list}

\pagebreak

\noindent{\bf FIGURE CAPTION}

\vspace{1cm}

\noindent FIG. 1. \ Behavior of the parameter $\, p\,$ as a function of
the root $\, x$.

\end{document}